# Wettability and "petal effect" of GaAs native oxides

A. Gocalinska[1,*], K. Gradkowski[1,2,†], V. Dimastrodonato[1], L. O. Mereni[1], G. Juska[1], G. Huyet[1,2], and E. Pelucchi[1]

[1] *Tyndall National Institute, "Lee Maltings", University College Cork, Cork, Ireland*

[2] *Centre for Advanced Photonics and Process Analysis, Cork Institute of Technology, Cork, Ireland*

## Abstract

We discuss unreported transitions of oxidised GaAs surfaces between (super)hydrophilic and hydrophobic states when stored in ambient conditions. Contact angles higher than 90° and high adhesive force were observed for several air-aged epitaxial samples grown under different conditions, as well as on epi-ready wafers. Regardless of the morphologies of the surface, superhydrophilicity of oxygen-plasma treated samples was observed, an effect disappearing with storage time. Reproducible hydrophobicity was likewise observed, as expected, after standard HCl surface etching. The relation between surface oxides and hydrophobic/hydrophilic behaviour is discussed.

*Corresponding author

†Current address: Institute of Chemistry, Hebrew University of Jerusalem, Givat Ram, Jerusalem 91904, Israel

Gallium arsenide-based semiconductor structures are widely used in optoelectronic, photonic and electronic devices and have been intensively studied for over 40 years now.[1] Together with these known device-related interests, a growing attention is also appearing for the exploitation of the well-established III-V technologies in the field of bio/medical applications, putting on them a requirement of working in aqueous conditions.[2] For example, epitaxial InAs quantum dots (QD) in GaAs matrix have been proposed as a platform for optical bio-sensing and devices.[3] For some of these applications a key issue is the understanding of the III-V and its oxides surface behavior, e.g. the wettability, their "exact" chemistry and the associated possibility of functionalizing them (all this obviously when a "normal" and appropriate storage is chosen). Moreover, unexpected physical properties have been reported for a number of semiconductor (and metal) oxides, showing non-trivial characteristic dependence on the detailed chemical configuration, in general opening for future applications.[4,5]

Air exposure of pristine GaAs forms layers of native oxide on its surface, containing various crystalline and amorphous forms of e.g. $Ga_2O_3$, $As_2O_3$ and $GaAsO_4$.[6,7] It is noteworthy that despite the known non-stoichiometry of moist air/water or even plasma-formed oxides[8], the wettability of the oxidized GaAs surface is commonly considered to be straightforward, while, surprisingly, only scant experimental data can be found in the literature. The generally diffused understanding is that "as grown" epitaxial GaAs is hydrophobic and the oxidized surface is hydrophilic.[9] One of the explanations proposed for this is that a high level of dangling bonds favors hydrophilicity, while more saturated bonds - hydrophobicity.[10]

From a pragmatic point of view, hydrophobicity and hydrophillicity can be distinguished according to a contact angle measured between the sample surface and a water droplet. A large contact angle (above 90°) reflects a hydrophobic surface while a low contact angle reflects a hydrophilic surface. Surface showing water drop contact angle (WDCA) < 5° (>150°) are referred to as superhydrophilic (superhydrophobic). The tilting angle of a solid surface when the droplet starts sliding downward is called the sliding angle and is relevant in case of many processing techniques, such as coating or cleaning, as liquids and solids, once attached, do not maintain a constant state. Both, contact and sliding angles are influenced by the micro- and nanoscaled morphology of the surface [11,12,13] and a proper design of the surface roughness can have strong impact on the solid-water contact area and its dynamics. What is more, recently a novel effect (hydrophobicity and high adhesive forces) was reported on planar, unpatterned surfaces of $HfO_2$[14], named "petal effect", resembling the phenomenon observed for the first time on rose petals.

We present in this paper a systematic study of oxidised (epitaxially grown), ambient stored, GaAs surfaces which show hydrophilic and hydrophobic behaviours, and which can be switched to superhydrophilicity by simple exposure to an oxygen plasma treatment. We discuss a broad range of structures resulting in various morphological features on the sample surface as well as data obtained on planar epi-ready GaAs wafers. The results open interesting technological perspectives for the exploitation of GaAs surfaces (e.g. our findings could have important impact in the field of heterogeneous wafer bonding) and call for an improved understanding of III-V surface chemistry.

All epitaxial samples here analyzed were grown by Metal Organic Vapour Phase Epitaxy (MOVPE) and their surfaces show morphological details with various arrangement of step flow/step bunching. This is typical of MOVPE processes, which involves decomposition and diffusion of precursor species and subsequent adatom diffusion and incorporation.[15,16] All growth runs for the structures described in this work were carried out at low pressure (20 mbar or 80 mbar) in a commercial horizontal reactor with purified $N_2$ as carrier gas. The structures, all capped with a GaAs layer, were grown on (001) GaAs perfectly oriented or slightly misoriented substrates.[17] The precursors were trimethylgallium (TMGa), trimethylaluminum (TMAl), trimethyindium (TMI) and arsine ($AsH_3$) or tertiarybutylarsine (TBA). Growth conditions and structural design varied from sample to sample, relevant details are referenced in the text when a particular example is discussed. As a reference we used various epitaxy-ready wafers on which the contact angle measurements were done without any initial surface processing or cleaning. The wafers were purchased from AXT, Wafer Technology and Sumitomo. It should be mentioned that these substrates were stored (in their original packaging) for longer time than producer guarantee (they were purchased more than 6 months before the experiment was conducted and stored in ambient conditions).

All epitaxial growths resulted in smooth, mirror-like surfaces, which were subsequently investigated with Atomic Force Microscopy (AFM) in tapping mode to provide detailed morphological information. WDCA measurements were conducted to determine hydrophilic/hydrophobic properties of the surface. One µl of deionised water (DI) was dispensed by micro-syringe on the sample surface and the contact angle of the formed sessile drop was measured. The profile of the droplet was recorded by a computer-controlled system and the contact angle was taken as the angle between the substrate surface and tangent to the droplet surface at the substrate/droplet/air interface. Multiple measurements were taken from a single sample, showing less than 2° deviation from the average value.

Measurements were conducted on "fresh" material (within 30 min after removal from the MOVPE reactor), and then repeated after 1 day, 7 days, 30 days and several months (or even years) of storage in ambient atmosphere. The oxygen plasma treatment was conducted in a Diener Electronic FEMTO Plasma System, at 50W and at 0.2 mbar for the time specified in the text. Wet chemical etching was performed by dipping the sample into 37% HCl aqueous solution and then rinsing with DI water. Samples were stored in a variety of standard laboratory carriers/shippers, like Fluoroware (polypropylene) carriers, as well as in transparent (poly)styrene and even in membrane carriers, with the surface never directly in contact with the carrier itself. We will discuss at the end of our contribution, that no appreciable differences were detected with storage carrier, and its relevance to unintentional contaminant artefacts. Nevertheless, we anticipate and stress here that our work is anyway relevant to standard storage and laboratory (cleanroom) practice, and, as such, relevant to a very broad scientific/technological community.

In Figure 1 we show AFM images of the surface profiles of two significantly different samples: the left panel refers to a planar 100 nm thick GaAs layer grown on 2° misoriented substrate (referred to as *planar GaAs*), on the right panel an example of GaAs cap covering a complex InAs QD structure is illustrated (referred to as *QD GaAs*, details regarding this material were discussed in Ref. 16). The planar sample shows a standard, for MOVPE, step bunched surface.[15] The *QD GaAs* sample surface on the other hand is covered homogeneously by elongated islands, on average 1x2 µm in lateral dimensions. The modulation in height, following a periodic pattern of apexes and notches, stayed within 25 nm range for individual feature. Crystallographic steps were clearly visible in both cases, confirming the epitaxial growth.[16]

In Figures 2 and 3 we present how the silhouette of the water droplet dispensed on the sample surface changed with the storage time in air and after treatment on the surfaces of planar epitaxial GaAs (left panel) and *QD GaAs* (right panel). The WDCA increased (actually irrespective of substrate misorientation choice for the epitaxially planar structures as checked with other samples) with time from below 25° ("fresh" material, Fig. 2a) to higher values (Fig. 2b and c), and for samples aged for several months the WDCA exceeded 90° (Fig. 2d). It must be said that both the *planar* and *QD* epitaxial samples initially showed rapid increase in the contact angle and eventually may breech the hydrophobic threshold. The only relevant difference is that while the *GaAs QD* samples show hydrophobic behaviours after a few months of aging, the planar epitaxial structures seems to take longer, reaching hydrophobicity only several months later, which suggests that perhaps the *QD GaAs* corrugated morphology accelerates the hydrophobicity process. All this is an indication that it is the growing oxide layer (grown in ambient conditions) that increased the contact angle in these structures, giving a promising possibility to reproducibly attain hydrophobic GaAs (oxide) surfaces (see Fig. 4 for a time dependence and a summary of our results).

It is noteworthy that the droplets on all aged samples showed a high adhesion, not sliding off when the sample was tilted, staying on the material surface even if the sample was turned up-side-down, similar to the rose petal effect case[14]. It should be said that such behaviour was observed in all investigated GaAs samples (and not only in the case of the two examples shown), regardless of the growth conditions (and misorientation of the substrate) and the exact details of the surface roughness, which can vary significantly depending on the design. The epi-ready substrates (as-bought from manufacturer), in this respect, behaved identically to the aged epitaxial samples (for more details see Fig 5 a and b later in the text).

To finally investigate the possibility of forcing the oxidation process, we treated several pieces of samples and substrates with oxygen plasma for 30 sec to 5 minutes. This resulted in an unexpected reduction of the WDCA to 0 (Fig. 3a), regardless of the plasma oxidation time. The superhydrophilic effect disappeared after a few hours when the samples were subsequently stored in ambient conditions and high contact angles were shown on all the test pieces (Fig. 3b), eventually reaching values similar to those of the air-only aged samples. The oxygen plasma did not perturb significantly the surface morphologies of all the samples investigated, with only the appearance of a number of expected small oxide pits as measured by AFM. We emphasize that the rapid recovery of the hydrophobic character of oxygen-plasma treated samples seems to indicate a temporary modification of the surface chemistry (possibly through the formation of hydroxyls[18]) which, in this case, quickly decays in the following hours at ambient conditions.

For completeness, we immersed fragments of the samples and substrates in 37% aqueous solution of HCl (which is a standard routine to remove most of the GaAs oxides) for a time from 5 to 15 minutes and after rinsing with DI water and blow-drying with nitrogen, we measured the WDCA again. The measured values of WDCA were increasing with treatment time, saturating after about 10 min (times varied for different samples) and reaching finally 80 - 100° (Fig. 3c).

We want to stress that, as all the samples were measured in ambient conditions, it is then impossible for them to stay uncovered without at least a thin layer of oxide (even just-grown or HCl-treated surfaces were exposed to air for several minutes). The contact angles measured again after storing the acid-treated sample in air for several days, showed the same values as just after treatment, showing clearly that it is not an oxygen-free surface the origin of the hydrophobic behaviour. It is an indication that perhaps it is the specific thin oxide layer formation after acid treatment and subsequent air exposure which is partly responsible of the hydrophobic behaviour.

To test the reproducibility of the plasma oxidation and HCl etching, we made several subsequent treatments interchanging those techniques. Regardless on sample history, the obtained WDCA was consistent with the value characteristic to the last treatment method (samples plasma-oxidised – HCl treated – plasma oxidized were superhydrophillic instantly with WDCA increasing with storage time; samples HCl treated - plasma oxidized - HCl treated were showing high WDCA and so on, as shown in Fig 5 c and d).

Before concluding our contribution, we need as well to discuss briefly the role of possible unintentional major contamination of our surface, causing, artificially, the (aging) effects we observe. One has to keep in mind that all carriers are potentially contaminating the samples: for example the "clean" Fluoroware carriers are known to degas over a long time period. These are mostly water and other ambient gases, which are obviously not an issue in this context, but also other organic compounds (including trimethylsilanol) are in the picture, as well as a number of metals and inorganic compounds, all not necessarily to be found on the sample surfaces. Over a long period of time some of those will be incorporated with the GaAs oxide which is growing on the sample surface. If the contamination process is slow in comparison with the oxide kinetic, all these will act as minor impurity inside the GaAs oxide matrix. Although it would be theoretically possible that some form of "greasy material" (or the like) has coated uniformly all the analyzed samples due to improper storage, hiding the real surface properties, in this particular case it appears to be very unlikely.

We intentionally utilized a variety of standard laboratory carriers, with no appreciable differences; differences which one would expect since all would contaminate the surface in a different way. It is also known that substantial contamination effects are observed over long period of time (a year is an appropriate unit for this, see for example Ref. 19), and what we observe saturates in less than three months in many samples, and evolution ceases from then on. We observe that this consideration is reinforced by the fact that it is known that industrial GaAs wafer suppliers guarantee their "epi-ready" surfaces for more than six months, making rather unlikely that over such a period a substantial contamination from the carriers to be an issue. It should also be observed that in the case of the oxygen-plasma-induced hydrophilicity we should assume that the surface changes (the surface would oxidize more and in a disordered way, incorporating some contaminants and if an organic contaminant would be present it would get removed in some way) and then manages to get contaminated exactly in the original way again in a few hours. Another reason stems from the fact that we intentionally degreased some test samples with standard acetone and IPA solutions. After DI water rinsing the surface properties went back to whatever they were before the procedure, excluding the presence of inorganic contamination of the surface (at least of those which are soluble in those solvents). Finally, we observe a faster hydrophobicity process on mesotextured surfaces. If the contamination was a major factor, one would expect differences in the surface organization to have a minor role.

In conclusion we have shown that assuming the presence of oxide on GaAs surface based simply on observation of the contact angle is incorrect. Oxidized surfaces (obtained in ambient conditions) show WDCA in broad range, even exceeding 90°. In particular we showed that mesostructured GaAs epitaxial samples can accelerate the hydrophobic process. The physical origin of this is unclear, and has to rely on the exact, subtle details of the surface chemistry of III-V surface oxides and the normally-adsorbed contaminants as a result of air/ambient exposure. Future high-resolution photoemission studies[20] might be useful to help clarify this point. Nevertheless the variety of effects we observe is solid and reproducible, rules out artifacts from major contaminants, and is to be observed in any modern laboratory which uses standard storage facilities. Moreover, the use of oxygen plasma and HCl etching seems to be reliable method for assuring, respectively, hydrophilic and hydrophobic behaviour for processing and fabrication purposes, opening interesting new technological perspectives.

This research was enabled by the Irish Higher Education Authority Program for Research in Third Level Institutions (2007-2011) via the INSPIRE programme, and by Science Foundation Ireland under grants 05/IN.1/I25 and 07/SRC/I1173. The authors are grateful to Dr. C. Colinge for useful discussions and to Dr K. Thomas for the MOVPE system support.

## References

[1] M. D. Hill and D. B. Holt, Journal Of Materials Science **3,** 244 (1968); Henry Kelly, Science **199**, 634 (1978)

[2] T. O'Sullivan, E. A. Munro, N. Parashurama, C. Conca, S. S. Gambhir, J. S. Harris, and O. Levi, Optics Express 18, 12513 (2010).

[3] Jan J. Dubowski, Lasers and Electro-Optics Society, LEOS 2006. Proceedings of 19th Annual Meeting of the IEEE (2006).

[4] H. Cui, G. Z. Yang, Y. Sun, and C. X. Wang, Appl. Phys. Lett **97**, 183112 (2010) and references therein.

[5] L. Chen, G. Henein, and J. A. Liddle, Proceedings of Nanotech 2009, 3, 195, Texas, USA, 2009.

[6] M. G. Proietti, J. Garcia, J. Chaboy, F. MorierGenoud, and D. Martin, J. Phys. Condensed Matter **5**, 1229 (1993).

[7] N. T. Barrett, G. N. Greaves, S. Pizzini, and K. J. Roberts, Surf. Sci. **227**, 337 (1990).

[8] Sidney I Ingrey, Surface processing of III-V Semiconductors, in *Handbook of Compound Semiconductors*, edited by: Paul H. Holloway, G. E. McGuire, Noyes Publications 1996 and reference therein.

[9] Thomas Mirandi and Douglas J. Carlson, CS MANTECH Conference, May 18th-21st, 2009, Tampa, Florida, USA.

[10] K. Matsushita, T. Monbara, K. Nakayama, H. Naganuma, S. Okuyama, and K. Okuyama, Electronics and Communications in Japan, **J-84C**, 134, (2001) and references therein.

[11] R. N. Wenzel: Ind. Eng. Chem. **28,** 988 (1936).

[12] A**.** B. D. Cassie and S. Baxter: Trans. Faraday SOC. **40,** 546 (1944).

[13] Z. Yoshimitsu, A. Nakajima, T. Watanabe, and K. Hashimoto, *Langmuir 18,* 5818 (2002).

[14] A. Tonosaki and T. Nishide, App. Phys. Express, **3,** 125801 (2010); L. Feng, Y.A. Zhang, J.M. Xi, Y. Zhu, N. Wang, L. Jiang, Langmuir **24**, 4114 (2008).

[15] E. Pelucchi, N. Moret, B. Dwir, D. Y. Oberli, A. Rudra, N. Gogneau, A. Kumar, E. Kapon, E. Levy, and A. Palevski, J. Appl. Phys. **99**, 093515 (2006); A. L.-S. Chua, E. Pelucchi, A. Rudra, B. Dwir, and E. Kapon, A. Zangwill, D. D. Vvedensky, Appl. Phys. Lett. **92**, 013117 (2008).

[16] K. Gradkowski, T. C. Sadler, L. O. Mereni, V. Dimastrodonato, P. J. Parbrook, G. Huyet, and E. Pelucchi, Appl. Phys. Lett **97**, 191106 (2010).

[17] V. Dimastrodonato, L.O. Mereni, R. J. Young and E. Pelucchi, Journal Crystal Growth 312, 3057 (2010).

[18] A. Sanz-Velasco, P. Amirfeiz, S. Bengtsson, and C. Colinge, Journal of The Electrochemical Society, **150,** 155 (2003).

[19] M. P. Seah and S. J. Spencer, J. Vac. Sci. Technol. A **21**, 345 (2003).

[20] B. Brennana and G. Hughes, J. of Appl. Phys. **108**, 053516 (2010).

**Figure 1.**

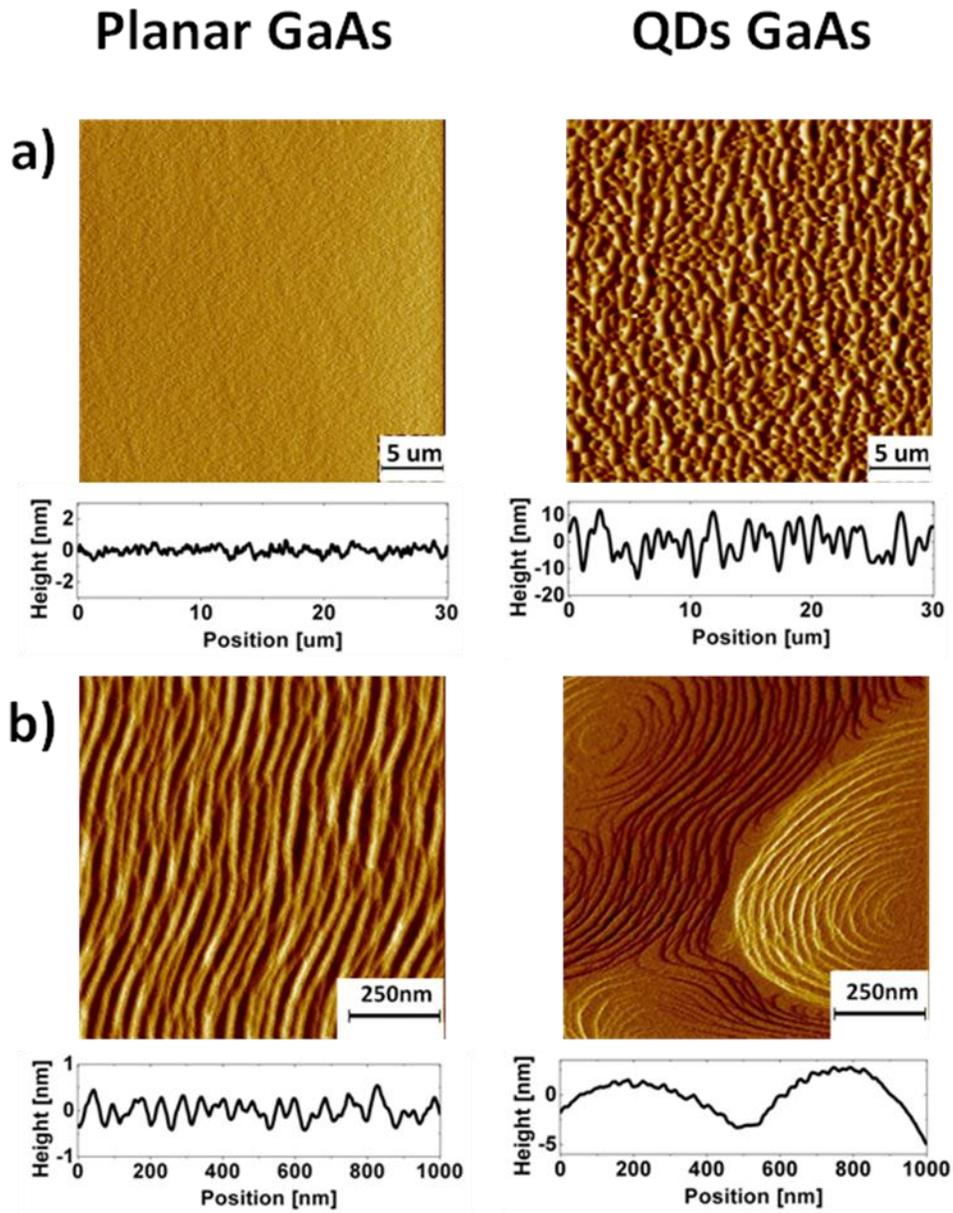

**Figure 2.**

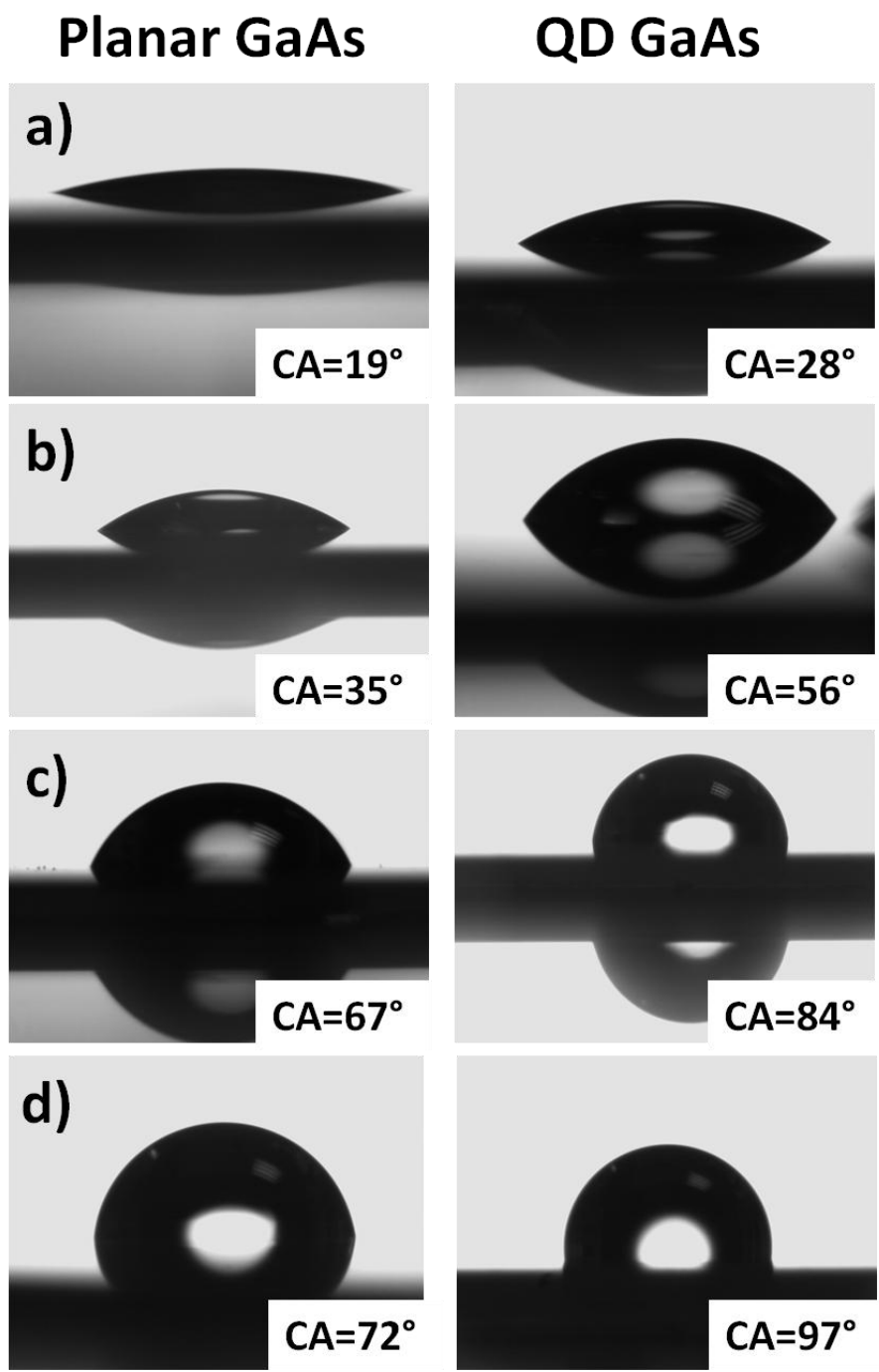

**Figure 3.**

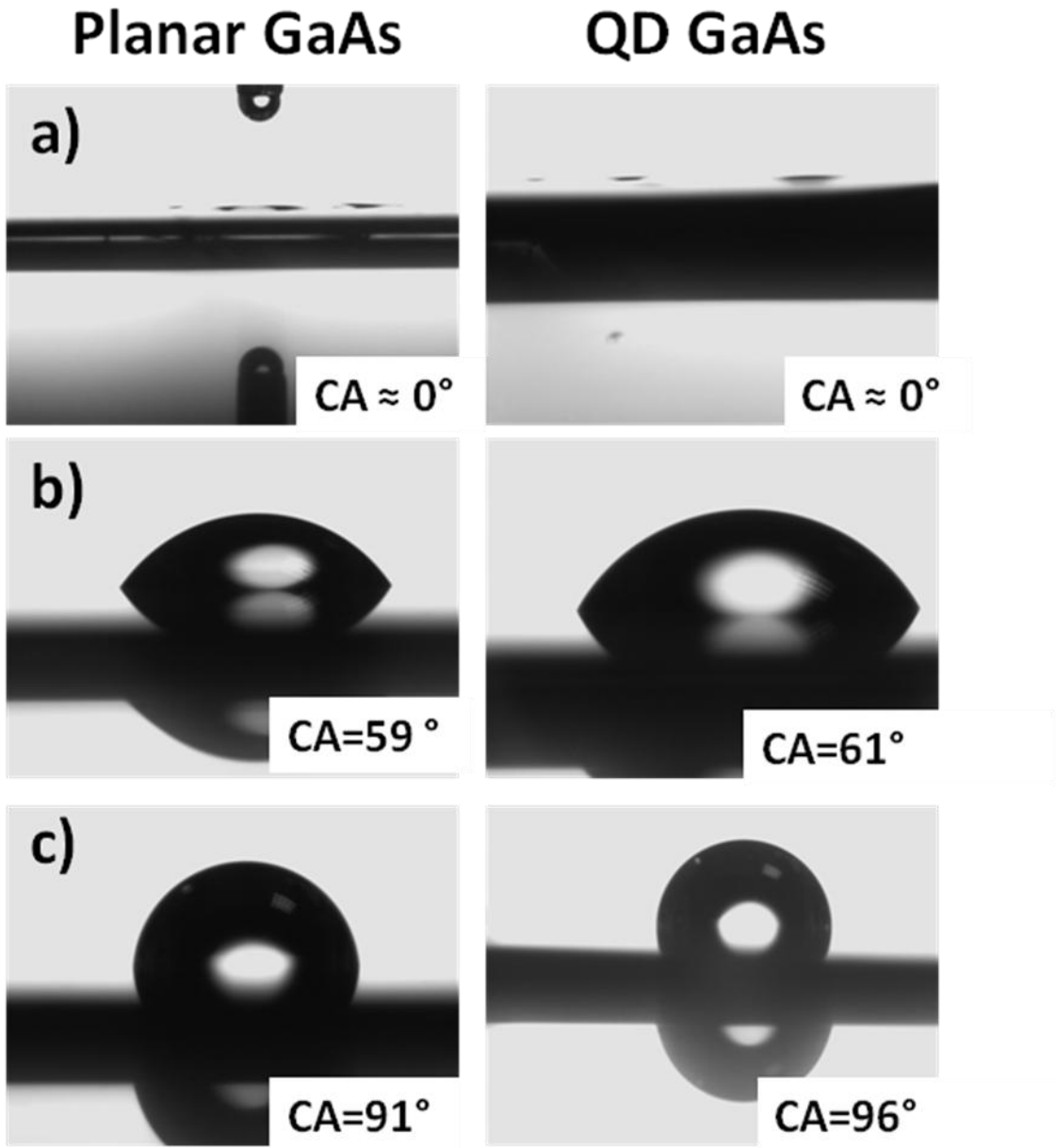

**Figure 4.**

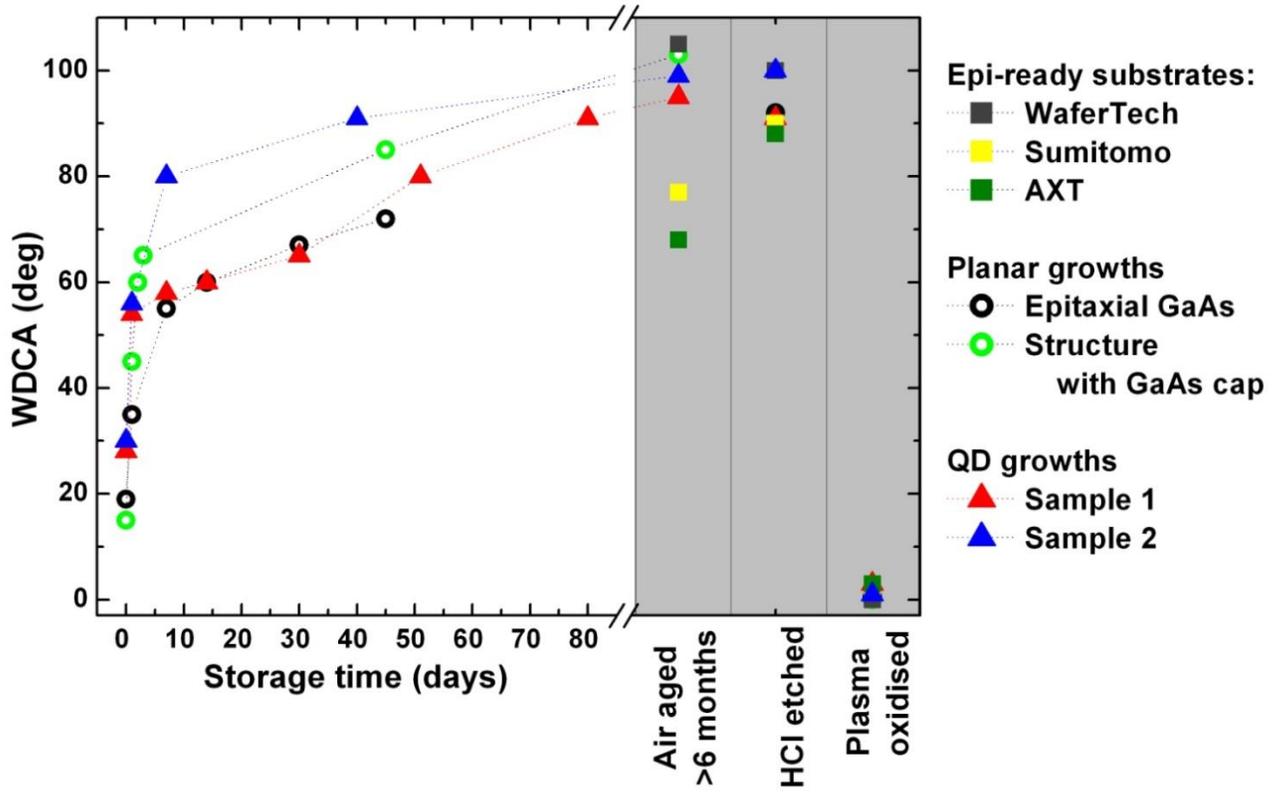

**Figure 5.**

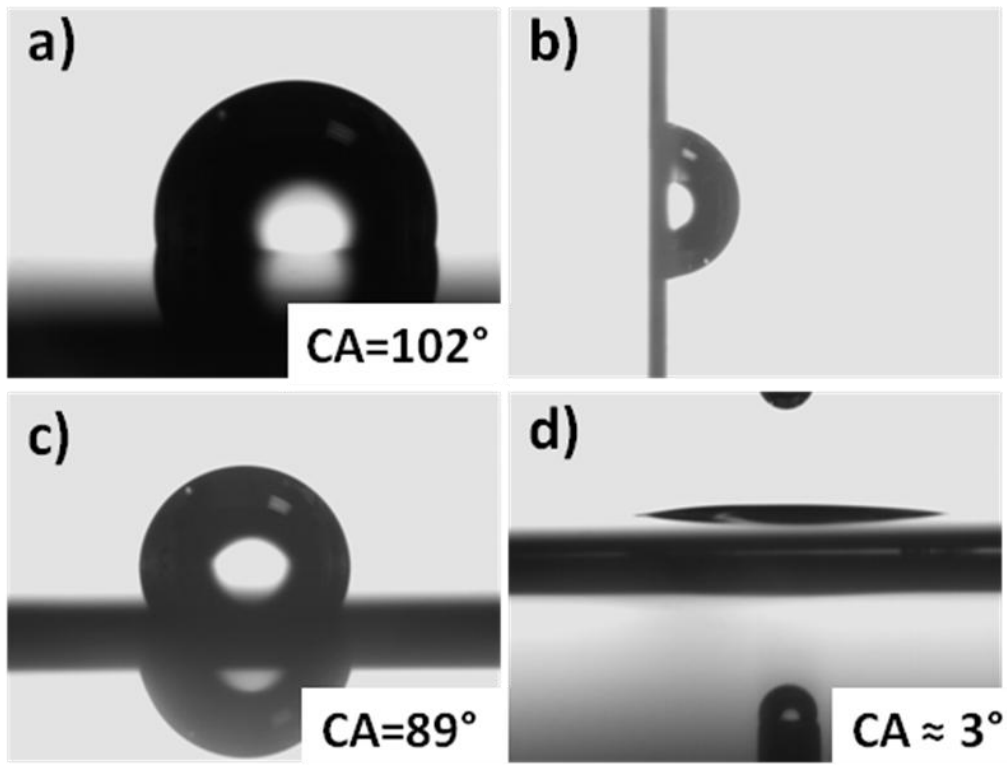

# Figure captions

**Figure 1.** (Color online) AFM image (signal amplitudes) of the top surface of the investigated structures, Left: planar GaAs, right: *QD GaAs* structure: a) large scale organization of the surface (30x30 μm), b) zoom-in to 1x1 μm area. Bottom panels show cross-section through the corresponding height images.

**Figure 2.** Photographs of water droplet silhouette on top of epitaxial structures, left panel corresponds to planar GaAs (epitaxial), right to *QD GaAs*; a) sample within 30min from removal from MOVPE reactor, b) sample air-stored for 7 days, c) sample air-stored for about 1 month, d) sample air-stored for several months.

**Figure 3.** Photographs of water droplet silhouette on top of epitaxial structures, left panel corresponds to planar GaAs, right to *QD GaAs*; a) sample after oxygen plasma treatment, b) sample after oxygen plasma treatment and subsequent storage for 1 day, c) sample after HCl etching

**Figure 4.** (Color online) Summary of WDCA measured on air-aged samples and after treatment. Measurement series are corresponding to individual samples with different growth conditions and morphologies. The square markers correspond to epi-ready wafers, circular to planar growths and triangular to quantum dot samples. The point corresponding to an epitaxially grown structure with a GaAs cap reported in the >6 month aged session corresponds to ~ 3 years of aging. Data points correspond to the average value measured on multiple drops deposited on the sample surface.

**Figure 5.** WDCA on epi-ready wafer by WaferTech; a) and b) as taken from the box, c) after plasma oxidation and subsequent HCl treatment, d) after HCL etching and subsequent plasma oxidation.